# A Fiber Optoacoustic Emitter with Maximized Conversion Efficiency and Controlled Ultrasound Frequency for Biomedical Applications

Linli Shi[1], Ying Jiang[2], Yi Zhang[3], Lu Lan[2], Yimin Huang[1], Ji-Xin Cheng[2,4*], Chen Yang[1, 4*]

[1] Department of Chemistry, Boston University, 580 Commonwealth Avenue, Boston, MA 02215, USA

[2] Department of Biomedical Engineering, Boston University, 44 Cummington Mall, Boston, MA 02215, USA

[3] Department of Physics, Boston University, 590 Commonwealth Avenue, Boston, MA 02215, USA

[4] Department of Electrical and Computer Engineering, 8 St. Mary's Street, Boston, MA 02215, USA

*corresponding author cheyang@bu.edu  jxcheng@bu.edu

**Abstract**

Focused ultrasound has attracted great attention in minimally invasive therapy, gene delivery, brain stimulation, etc. Frequency below 1 MHz has been identified preferable for high-efficacy drug delivery, gene transfection and neurostimulation due to minimized tissue heating and cell fragmentation. However, the poor spatial resolution of several millimeters and the large device diameter of ~25 mm of current sub-MHz ultrasound technology severely hinders its further applications for effective, precise, safe and wearable biomedical studies and clinical use. To address this issue, we report the development of a novel fiber-based optoacoustic emitter (FOE). The FOE, a new miniaturized ultrasound source, is composed of an optical diffusion coating layer and an expansion coating layer at an optical fiber distal end with a diameter of approximately 500 microns. Taking advantage of the fiber size and diffusive nanoparticles introduced, the ultrasound generated by the FOEs showed a spatial confinement of sub-millimeter. The optoacoustic conversion efficiency was maximized through choosing absorbing nanomaterials and thermal expansion matrix. Controllable frequencies in the range of 0.083 MHz to 5.500 MHz were achieved through using the diffusion layer as a damping material or modifying the nano-composition in the expansion layer. This sub-MHz frequency controllability allows FOEs to be used as a localized ultrasound source for precise cell modulation. We demonstrated optoacoustic cell membrane sonoporation with a localization of sub-millimeter and negligible heat deposition, implicating its broad biomedical applications, including region-specific drug delivery, gene transfection as well as localized neuron stimulation.

**Introduction**

The past decades have seen a rapid development of focused ultrasound technologies for noninvasive or minimally invasive cellular biotechnology, such as drug delivery and gene transfection [1-4]. Ultrasound mediated sonoporation has been extensively used for treatment of tumors with dense stroma such as pancreatic tumors, showing introduction of less systemic toxicity through less circulating drug required than traditional chemotherapy [5]. Focused ultrasound mediated gene therapy and neuron stimulation have also received considerable attention for treating immunodeficiency disorders, Parkinson's disease and certain types of cancer [6,7]. Ultrasound of high frequencies (above 1 MHz) carries great risk of tissue heating [8] and cell fragmentation [9]. Meanwhile, low frequency ultrasound was shown to be more effective. 0.2 MHz ultrasound induces 7 times lower threshold in cavitation compared to high frequency (4.8 MHz) [10], and ultrasound of 0.3 MHz shows 150 times lower pressure threshold for successful neuron stimulation compared to 3 MHz [11]. Thus, frequency within the range from 20 kHz to 1.0 MHz is considered to be preferable in biomedical applications [12-14], including drug delivery [2,15], gene transfection [16,17], neuron stimulation [11] and blood brain barrier disruption [4].

However, the conventional sub-MHz frequency ultrasound transducers are bulky and poor focusing. A typical piezo transducer producing ultrasound with a frequency of 1.06 MHz comes with a diameter of 25 mm [18]. Efforts have been made to fabricate miniaturized transducers, including low-frequency flex tensional resonators [19], tonpilz transducers [20], and thickness-type resonators [21]. For example, aiming at ultrasound mediated drug delivery, a piezoelectric disc with an unprecedented thickness of 1 mm and a diameter of 12.7 mm has been developed to provide acoustic between 1 kHz and 100 kHz depending on the geometry [22]. Nevertheless fabrication of these transducers with millimeter-scale lateral dimensions is considered challenging and expensive[23]. In addition to its large device size, the transducer-based focus ultrasound technology produces a large diffraction limited focal volume at millimeter scale for an ultrasound of a few hundred kHz. The ultrasound wave of 1 MHz generated by the transducer was found to have a lateral diameter of 6 mm[24]. Since the diffraction limit is reversely proportional to the frequency, a desired frequency of 0.2 MHz ultrasound has a larger focusing diameter with a size similar to a whole mouse brain (~5.5×8×14 mm) [25], making it impossible to pinpoint a specific region of the brain. In the study of brain stimulation and gene transfection, a sub-millimeter spatial resolution will be beneficial. For example, the subthalamic nucleus (STN) is the targeted area of gene transfection for Parkinson's disease[26]. The stimulation of sub-territories of the STN revealed its role in the integration of the emotional and motor aspects of behavior[27]. Thus, sub-millimeter resolution is required to study the function of these subterritories in animal models, which is beyond the capability of traditional transducers.

Optoacoustic, where pulsed excitation light is absorbed by materials of interest, resulting in transient heating, material compression and expansion, and subsequently pressure change, is a novel way to generate ultrasound. It was shown as an emerging technique for biomedical imaging[28]. Fiber-based optoacoustic emission has been explored for miniaturization taking advantage of submillimeter diameters of optical fibers. Thus far, research of fiber-based optoacoustic generation was mainly focused on imaging, targeting an acoustic frequency with a wide bandwidth of tens of MHz[29]. Specifically, in Colchester's work, carbon nanotubes (CNTs) were mixed in polydimethylsiloxane (PDMS), followed by dip coating on an optical fiber tip, which led to a peak frequency at 18 MHz and a bandwidth of 12 MHz[23]. Similarly, Noimark et al.[30] and Poduval et al.[31] coated CNTs and subsequently PDMS on the tip of an optical fiber and showed peak frequency of 20 MHz and 30 MHz, with bandwidth of 23~40 MHz and 29 MHz, respectively.

Notably, none of reported fiber based optoacoustic devices in the literature delivered sub MHz frequency, which is in the need of cell modulation with high efficiency and minimized safety issue. Thus, these limitations of current ultrasound and optoacoustic technologies highlight an unmet need of a novel miniaturized low-frequency acoustic source with controllable characteristics and submillimeter spatial precision. Such a device will enable precise and effective cell modulation for targeted therapies, and open up potentials for broader biomedical applications when integrated with other medical devices.

In the present work, we report a fiber-based optoacoustic emitter (FOE) with a controllable frequency spectrum, including peak frequency, targeting the low frequency range of sub-MHz. A key innovation of our device is to design and coat the fiber tip with two distinct functional layers: an optical diffusion layer and a thermal expansion layer, to deliver sufficient ultrasound intensity and to control the peak frequency and bandwidth needed for cell modulation. In addition, we investigated the effect on photoacoustic intensity by varying materials for the coating to achieve maximized intensity needed to biomedical applications. Employing the FOE that delivers sub-millimeter high special precision ultrasound with optimized frequency and intensity, we demonstrated effective delivery of small molecules into living cells via sonoporation effect with less than 1 ˚C temperature increase. Our work offers a new ultrasound technology for cell modulation applications.

**Material and Methods**

**Design and fabrication of a two-layer fiber-based optoacoustic emitter**

The basic design of the FOE is schematically represented in **Figure 1**. To achieve miniaturization, optical fibers were utilized for the laser transmission (Fig.1b). The FOE is constituted by a light diffusion layer and an absorption/thermal expansion layer (Fig.1c and d).

The diffusion layer was introduced to prevent localized heating and subsequent damage of the expansion layer due to the difference of thermally induced strain within the layer[29]. This diffusion layer comprises a mixture of polymer (Epoxy or PDMS) and 100-nm diameter zinc oxide (ZnO) nanoparticles, which diffuse the high-energy laser pulse into a Cauchy distribution owing to its high optical transparency in the near infrared region and high refractive index [32]. The diameter of ZnO nanoparticles (i.e. 100 nm) is much smaller than the laser wavelength (1030 nm) used, enabling Raleigh scattering in all directions[33]. Then, to convert the light energy into acoustic waves, an absorption/thermal expansion layer was subsequently added as the second coating. It is composed of nanoparticles with high light absorption coefficient as the absorber (graphite or multi-wall carbon nanotubes, MWCNTs) and polymer with high thermal expansion coefficient for the purpose of expansion and compression (Epoxy or PDMS). Taking advantage of these specially designed nano-polymer composite layers at the fiber distal end, upon the pulsed laser excitation, an acoustic wave was effectively generated from the fiber tip through the optoacoustic effect.

In order to achieve high optoacoustic conversion efficiency, materials with superior physical properties were chosen. According to the basic principle of optoacoustic wave generation[28,29,34], for a given material, the initial optoacoustic pressure is $P_0 = \xi \Gamma \mu_a F$, where $\xi$ is the heat conversion efficiency, $\mu_a$ is the optical absorption coefficient and $F$ stands for the optical fluence. The Grüneisen parameter $\Gamma$ is defined as: $\Gamma = \beta v_s^2 / C_p$, where $\beta$ is the isobaric volume thermal expansion coefficient, $v_s$ is the speed of sound, and $C_p$ is the specific heat capacity at a constant pressure[28,35]. Thus, material with large light absorption coefficient $\mu_a$ and thermal expansion coefficient $\beta$ enables more absorption and subsequently converted into stronger acoustic waves. A preferred choice is the combination of MWCNTs and PDMS. MWCNTs are known for high light absorption coefficient (0.45 μm-1) compared to graphite (4~18 μm-1)[36]. PDMS has a high volumetric thermal expansion coefficient ($\beta \approx 960$ μm/m·˚C) [37], which is 5-7 times higher than the epoxy ($\beta \approx 132\text{-}192$ μm/m·˚C)[38]. For this reason, the mixture of MWCNTs and PDMS was tested for coating on the fiber distal end. To further elucidating the efficacy improvement based on matrix properties, the mixture of epoxy and graphite was used for comparison.

The novelty of FOEs with the two-layer coating design and the material choice can be highlighted by comparing it to the typical design of a transducer. A transducer is composed of three layers: a matching layer to match the acoustic impedance between medium and the active layer; an active layer (piezo-electric materials to generate ultrasound upon applied voltage), and a backing layer to match the acoustic impedance between the active layer and the back connector. Since the acoustic impedance of PDMS is close to water (1.48 Pa·s/m3), the matching layer can be spared in the FOE. The frequency is reversely proportional to the thickness of the active layer. The damping effect of the backing layer also controls the frequency and bandwidth [39]. In the two-layer FOE design, the epoxy (acoustic impedance: 2.5-3.5 Pa·s/m3 ) in the diffusion layer acts as the backing layer matrix to match the acoustic impedance between the fiber (silica, acoustic impedance:13.1 Pa·s/m3) and the active layer (PDMS, acoustic impedance: 1.1-1.5 Pa·s/m3)[40,41].

The fabrication has two steps. First, for the diffusion layer, the Epoxy or PDMS matrix was prepared via cross linking process. Epoxy was made by mixing polyepoxides solution (Devcon Inc, Alberta, Canada) with polyfunctional curatives in a radio of 1:1 by volume. For PDMS, the silicone elastomer (*Sylgard* 184, Dow Corning Corporation, USA) was dispensed directly into the container carefully to minimize air entrapment, followed by mixing with the curing agent in a ratio of 10:1 by weight. Subsequently, ZnO nanoparticles serving as diffuser (~100 nm, Sigma-Aldrich, Inc., MO, USA) were added into the matrix at a concentration of 15% by weight otherwise specified. The concentration was chosen based on previous optimization study to achieve a uniform distributed laser emission[33]. A multimode optical fiber with 200 µm core diameter (FT200EMT, Thorlabs, Inc., NJ, USA) and a polished distal end was carefully dipped about 100 µm below the surface of the mixture solution and then quickly pulled up, using a micro manipulator. By placing vertically at room temperature, the polymer crosslinked and the matrix formed the coating. The diffusion layer made of Epoxy was subsequently coated with the absorption/thermal expansion layer of Epoxy. In this way, the acoustic impedance mismatch was minimized, providing the maximized optoacoustic conversion efficiency. Graphite powder (Dick Blick Holdings, Inc., IL, USA) was mixed with the matrix at a concentration of 30% by weight. MWCNTs, (<8 nm OD, 2-5 nm ID, Length 0.5-2 µm, VWR, Inc., NY, USA) was used at a concentration of 0-10% by weight, approaching the solubility upper limit owing to its low density (1.65 g/cm[2]). Similarly, the FOEs made of PDMS matrix were fabricated. In the latter experiment of FOE generating the sub-MHz frequencies, the structure was modified as ZnO/Epoxy (the diffusion layer) and CNTs/PDMS (the absorption/thermal expansion layer) to realize acoustic impedance mismatch. In this way, the pressure was compromised while still meeting the frequency need for cell modulation. By making a mark near the fiber tip with thermal resist ink, the thickness after coating was measured by aligning the mark on the before-and-after micrograph.

Light leak would generate acoustic response on the detecting transducer due to the pulsed laser induced shockwave in water and transducer probe[42], and potentially cause damage of the detector. A power meter was used to measure light transmittance to evaluate the light leak. We found the design and fabrication discussed above eliminated the light leak of FOEs. For all FOEs with different materials and geometric structure throughout the study, a light transmittance of less than 0.5% was ensured by multiple coating of absorption/thermal expansion layer.

**Ultrasound generation and characterization**

A thorough acoustic characterization of the generated optoacoustic waves was carried out by a setup shown in **Figure S1**. A Q-Switched laser (Bright Solutions, Inc., AK, USA) with a wavelength of 1030 nm and a pulse width of 3 ns was used as the laser source. The laser pulse energy was 127 mJ/cm$_2$ unless otherwise specified. Two functional generators were used to give laser pulses in a tone burst mode. Functional generator 1 provided triggers with repetition rate of 0.5 Hz as the tone burst frequency. For each burst, the functional generator 2 provided triggers with repetition rate of 1.7 KHz and burst duration of 0.2 s. In the measurement setup, an ultrasound transducer (5 MHz, V326, Olympus, MA) together with an ultrasonic pre-amplifier (0.2–40 MHz, 40 dB gain, Model 5678, Olympus, USA) was utilized to characterize the frequency of the ultrasonic waves. To quantify the acoustic pressure, a needle hydrophone with a diameter of 40 μm and frequency range of 1-20 MHz (NH0040, Precision Acoustics Inc., Dorchester, UK) was utilized together with the pre-amplifier. The distance between the FOE tips and transducer was kept at 1.50 mm and at 0.1 mm between the FOE tips and hydrophone. A digital oscilloscope (DSO6014A, Agilent Technologies, CA) was used to display the readout electrical signal from the transducer or hydrophone. The FOE and transducer/hydrophone were all immersed in deionized water to mimic the typical environment related to biomedical application and to minimize the acoustic impedance mismatch among the FOE, water and detectors.

**Cell culture and fluorescence microscopy**

For the sonoporation experiments, MIA PaCa-2 (Human Caucasian pancreatic carcinoma cell, American Tissue Cell Culture Manassas, VA, USA) were seeded in 35 mm glass bottom dishes with a cell density of 8 ×10$_4$/ml and a confirmed viability of 80-95% prior to the ultrasound treatment. Sytox Green Nucleic Acid Stain (Thermo Fisher Scientific, Waltham, MA, USA) in phosphate buffer solution (PBS) was added to the cell culture dish to reach a final concentration of 10 μM immediately followed by the ultrasound treatment with conditions described later. To minimize the influence of laser induced cytotoxicity, time-lapse fluorescence images were taken every 2 min using an inverted wide-field fluorescence microscope with a LED at 470 nm as excitation light source. Images were acquired by a scientific CMOS camera (Zyla 5.5, Andor).

## Results and Discussion

### Maximizing optoacoustic conversion efficiency upon designing absorption/expansion layers

To evaluate the efficacy of CNTs/PDMS as an optoacoustic emitter for strong and narrow band frequency ultrasound, FOEs with absorption/expansion layers composed of Graphite/Epoxy, CNTs/Epoxy, Graphite/PDMS and CNTs/PDMS with the diffusion/expansion layer thicknesses of ~100 μm/~100 μm were fabricated. Notably, the current fabrication produce a coating thickness variation of ±8% assure identical coating thickness, the thickness variation induced amplitude fluctuation was observed to be minimal, demonstrated in **Figure S2**.

**Figure 2** shows the output waveforms from FOEs. Each curve was obtained by averaging 3 waveforms. The FOE with CNTs/PDMS exhibited a peak to peak pressure of 0.04 MPa (Green), whereas 0.0033 MPa, 0.0035 MPa and 0.0055 MPa for Graphite/Epoxy (black), CNTs/Epoxy (red) and Graphite/PDMS (blue), respectively. CNTs/PDMS increased the acoustic pressure by 12 folds compared to Graphite/Epoxy. This significant increase is attributed to the higher absorption of CNTs and thermal expansion coefficients of PDMS. According to the optoacoustic pressure equation, with a 9-fold increase in light absorption coefficient and a 5-fold increase of thermal expansion coefficient, the enhancement could reach up to 45 times. However, the concentration of CNTs was 10% whereas that of graphite was 30%, which could account for the one third reduction, leading to a final enhancement of 12 folds.

**Controlling the ultrasound frequency via modification of the diffusion layer**

The damping effect of the backing layer in a typical transducer impacts on the frequency produced, therefore we expect the change of the thickness of the diffusion layer, acting as the backing layer, will control the output frequency of FOE. We fabricated FOEs with ZnO/Epoxy diffusion layer thickness of 36, 42, 53, 62, 79, 100 μm, respectively. Then absorption/thermal expansion layers of CNTs/PDMS with a thickness of 109±24 μm were used. It's worth noting that the variation of the thermal expansion layer physical thickness in this experiment doesn't change the ultrasound frequency, which will be explained in the next section. The time-of-flight optoacoustic signals were recorded (**Figure 3a**), processed with FFT and shown in the frequency domain (**Figure 3b**). The peak frequency was shown to be controlled in the range of 0.384 to 0.088 MHz through varying the diffusion layer thickness from 36 to 100 μm, suggesting a significant decrease in frequency while increasing the diffusion layer thickness. By examining the frequency range of 0-10 MHz, the distribution of frequencies exhibited a clustering at sub-MHz region (**Figure 3c**). In addition, the peak frequency plotted as a function of diffusion layer thickness demonstrated a linear relationship (**Figure 3d**). Controlling the frequency through changing the diffusion layer can be further rationalized by the fact that the peak frequency of the optoacoustic spectrum could be modulated by the mass of the diffusion layer. The optoacoustic effect can be described by the thermal expansion equation, which is a derivative of the generalized Hooke's law and the equation of motion that is deduced from Newton's second law. During the optoacoustic conversion process, the FOE tip can be regarded as a harmonic oscillator, in which the oscillating motion comes from the initial force given by the thermal expansion effect. In Hooke's law, the amplitude of the oscillation remains constant, and its frequency is independent of its amplitude, but determined by the mass and the stiffness of the oscillator[43].

**Controlling the ultrasound frequency via altering the CNT concentration in the expansion layer**

Another strategy to control the frequency is to change the effective thickness of the absorption/thermal expansion layer. As shown in the optoacoustic generation theory[44-46], the waveform of optoacoustic is also depending on the light absorption profile of the optoacoustic source, which is $\tau + 1/c\alpha$ ($\tau$ is the laser pulse width, $\alpha$ is the light absorption coefficient of the absorber). The effective thickness of the absorber is a measure of the light penetration depth. Therefore, according to the photoacoustic theory, the optoacoustic signal waveform consequentially changes with the effective absorber thickness. When the physical absorber thickness is larger than the light penetration depth, the extra thickness only induces acoustic attenuation.

To verify this, firstly, we investigate how sensitive the change of the peak frequency is to the change of the physical thickness of expansion layers. Two groups of FOEs were fabricated with ZnO/Epoxy diffusion layers with 36 μm and 100 μm, respectively. For each group, 4 FOEs were made with CNT/PDMS expansion layers varied from 100 μm to 210 μm as indicated by the color legend in **Figure S2**. **Figure S2** shows the time domain of ultrasound from FOEs. The waveform remained as similar functions with respect to time while the amplitude was dropping when increasing the expansion layer thickness. These suggested that the frequency, determined by the time dependence of the wave forms, are not sensitive to the change of the physical thickness of the expansion layer in the tested range, while the amplitude changing could result from the acoustic attenuation by the extra thickness of the thermal expansion layer.

While the physical thickness is not a controlling factor for the frequency, we expect to vary the frequency by changing the effective absorber thickness. This can be achieved through modifying the spatial absorption profile of the expansion layer via changing the absorber concentration while maintaining the same physical thickness. Since the effective thickness is primarily determined by the light absorption profile but not the physical thickness, in this way the influence of fluctuation in the physical thickness and geometric structures would be minimized, improving the robustness of the FOE fabrication.

To verify how the absorber concentration can be modified to fine-tune the ultrasound frequency, FOEs were coated only with the CNTs/PDMS expansion layers without the diffusion layers. Different concentrations of 2.5%, 5.0%, 7.5%, 10.0% by weight, respectively, of CNTs were used in the mixture. The thickness of the overall coating was kept in the same range. We expect that mixture with lower CNT concentrations allows higher light penetration depths, which subsequently increases the effective thickness, according to the Beer–Lambert law. The results are shown in **Figure 4.** From the frequency domain in **Figure 4a)**, the FOE with the CNT concentration of 2.5 % generated acoustic waves with a peak frequency of 1.0 MHz compared to the 5.5 MHz from FOE of 10.0 % CNTs concentration. The peak frequency was observed to increase from 1.0 to 5.5 MHz when increasing the concentration. Such CNT concentration dependent frequency change suggests that the controllable peak frequency of optoacoustic can also be achieved by modifying the light absorption profile of the absorption/expansion layer, which is in-line with the acoustic theory discussed above.

Collectively, these two complementary assays (**Figure 3 and 4**) demonstrate that, by integrating the frequency control ability of both the diffusion layer and the thermal expansion layer, we can achieve fine tuning of the frequency within the sub-MHz range.

**Delivering controlled pressure suitable for cell modulation**

To deliver a wide range of pressure through the ultrasound generated, we varied laser pulse energy in the FOEs. Optical attenuators with different air gap lengths from 0.4 mm to 5.8 mm were used to vary laser pulse energy ranging from 12 to 127 mJ/cm$_2$ measured by a power meter. A FOE with 36 µm diffusion layer (ZnO/Epoxy, 15% by weight) and 80 µm absorption/expansion layer (CNTs/PDMS, 10% by weight) was tested. The acoustic waveforms generated were compared when decreasing pulse energy (**Figure 5a**). The peak-to-peak amplitudes of each acoustic signal were plotted as a function of the laser pulse energy and showed a linearly dependence (**Figure 5b**). This is consistent with the basic optoacoustic equation mentioned above where the amplitude of photoacoustic signals is expected to be proportional to input laser pulse energy. The peak frequency in each spectrum was found to be 0.34 MHz, independent of the laser pulse energy (**Figure 5c and d**), whereas the FWHM slightly increased from 0.25 MHz (at 127 mJ/cm$_2$) to 0.37 MHz (at 39 mJ/cm$_2$) (**Figure 5c**). Collectively, these results confirmed that the laser pulse energy changed the acoustic amplitude but not the ultrasound frequency, which allows for controlling the acoustic pressure independently while maintaining the frequency.

**Producing consistent frequency in all directions.**

We further characterized the angular distribution of the acoustic wave in terms of amplitude and frequency spectra. We used a FOE composed of a diffusion layer (ZnO/Epoxy, 40 µm thick) and an expansion layer (CNTs/PDMS, 120 µm thick). The acoustic radiation from the FOE was determined by measuring the output voltage on the oscilloscope at a constant light input of 127 mJ/cm$_2$. The angle of the transducer detector relative to the fiber axis was varied by a controllable 360° rotation stage (Thorlabs, Inc., NJ, USA) with an accuracy of ±1 degree. Optoacoustic signals were acquired at angles of 0˚, ±25˚, ±50˚ and ±75˚, respectively, as illustrated in **Figure 6a. Figure 6b and c** show the measured acoustic amplitudes. The peak to peak photoacoustic amplitude in **Figure 6c** was found to decrease from 4.3 (at 0°) to 1.9 and 2.3 (at +75°and -75˚), respectively, which indicated that the larger the angle, the weaker the acoustic amplitude it was and it showed the maximum amplitude at the front direction. **Figure 6d** shows that the peak frequency was relatively constant (0.7-0.8 MHz) when varying the angle from 0°to ±75˚. Since majority of the laser pulse energy was delivered along the fiber axis direction, despite the diffusion introduced by the diffusion layer, less light propagates laterally, making the lateral optoacoustic induced vibration less. Other factors, including the curvature radius of the spherical coating, the discrepancy of density and sound speed between PDMS and water, could also contribute to the amplitude anisotropy. In previous optoacoustic simulation study[47], the optoacoustic wave was conforming to the shape of the fiber tip, which was modeled as a Dirichlet boundary condition on the tip surface. The outer limits of the liquid domain, which was modeled as water, was implemented as plane wave radiation boundary condition. Our finding is consistent with these simulation results, confirming the acoustic wave was scattered and subsequently propagated in all directions, while the acoustic spectrum exhibiting isotropy in frequency domain.

**The low-frequency ultrasound from the FOE allows deep tissue localization.**

Deep tissue penetration benefiting from sub-MHz ultrasound is critical for biomedical applications. Another key feature of the FOE is the low frequency capability, which held the promise of serving as a guided wire for surgical purpose. To specifically assess the acoustic penetration depth of the low frequency and high amplitude acoustic signals enabled by the FOE, a stack of chicken breast and a mouse skull were used as two models to analyze the penetration capability as shown in **Figure 7**. A FOE was used to produce acoustic signals of 0.14 MHz. A transducer with peak frequency of 10 MHz in the transmit mode was used as an acoustic source for comparison. A transducer covering frequency of 0-10 MHz in the receive mode was used for the detection (**Figure 7a,c**). **Figure 7b** shows the acoustic amplitude attenuation with the increased depth in the chicken breast. Black and blue data points indicate the attenuation of acoustic signals generated by the FOE and the 10 MHz transducer, respectively, suggesting that the lower frequency of 0.14 MHz exhibited a penetration depth of 10 cm, higher than the 4 cm for acoustic generated by 10 MHz transducer. Similarly, **Figure 7d** shows the acoustic signals after penetration of the mouse skull with the thickness of 0.15 mm. The acoustic was only reduced to 94.2±2.6% for the FOE, while for 10 MHz transducer, 76.2±3.4% of the acoustic remained. These data demonstrated the advantage of low frequency acoustic generated by the FOE. The smaller attenuation of the low frequency ultrasound generated by the FOE allows a deeper penetration depth in highly scattered biological tissues, such as brain.  Collectively, we demonstrated a miniaturized fiber-based FOE for localized acoustic emission with controllable low frequency in the range of 0.08-5.50 MHz and tunable amplitude suitable for cell modulation. The FOE also exhibits frequency independent of detection angle and significant penetration across 10 cm chicken breast and a rat skull, which was potentially capable of localizing medical subject such as a breast tumor deep in the tissue, thus replacing the current guide wire.

**The low-frequency ultrasound from the FOE enables regional Sytox cellular uptake within an area of 0.2 mm$_2$.**

To demonstrate the potential of FOE as a new device producing localized ultrasound with low frequency to modulate biological systems, we show that FOE directly enables cellular uptake of membrane-impermeable small molecules. Specifically, we chose to test the cellular uptake under sonoporation using a high-affinity intercalating nucleic acid stain-SYTOX Green, which only penetrates into cells through a compromised plasma membrane and displays fluorescence enhancement upon binding to nucleic acids. A FOE with a peak frequency of 1.4 MHz was placed approximately 100 µm above the focus plane of the fluorescence microscope in the center of the field of view. Ultrasound bursts of 200 ms duration were generated using a pulsed laser with 39 mJ/cm$_2$ at a 1.7 kHz pulse repetition rate, which corresponded to approximately 340 acoustic pulses per burst (**Figure 8a**). The optoacoustic treatment was performed at a burst repetition rate of 0.5 Hz in a total period of 10 minutes.

To exclude the thermal induced cell membrane permeabilization, we measured temperature increase at the fiber tip during the FOE treatment using a miniaturized ultrafast thermal probe (20 µm in diameter) placed in contact with the fiber tip. The temperature rise was found to be 0.6 °C within a total duration of 8 min (**Figure 8b**). At such small temperature increase, thermal-induced membrane depolarization is negligible. Therefore, the Sytox uptake results are attributed to mechanical disruption induced by the optoacoustic wave from the FOE. **Figure 8c** shows the fluorescence images of the cell culture before and after the FOE treatment compared with the control group with Sytox only and no ultrasound treatment. It's clear that when cells are treated with the FOE cellular uptake of Sytox is significantly increased, indicated by the increase in fluorescence signals. The fluorescence signals from 30 individual cells were taken and the average fluorescence intensity was plotted as a function of time in **Figure 8d**. The average signal was found to increase by 28% within the first 10 min when FOE was on, followed by continuous increasing to 70% at 26 min, indicating diffusing of Sytox through cell membrane to nuclei and bound to the DNA. The dynamic is consistent with the previous study reported, in which focused ultrasound facilitated the Sytox uptake on the time scale of tens of minutes[48]. To validate that the FOE provides a unique strategy to enable regional cell modulation through the localized delivery of specific molecules to cells by the confinement of the sonoporation, compared to typical whole cell dish modulation, fluorescence images were taken with a 10× field. The localized delivery was shown in **Figure 8e-f**. After the FOE treatment, the treated region with an area of 0.2 mm$_2$ exhibited significant fluorescence change of 32%. Furthermore, using 2 μg/mL Propidium Iodide staining ((Thermo Fisher Scientific, Waltham, MA, USA), the cell viability after FOE treatment was $99.55 \pm 0.03\%$ (Figure S3), indicating the superior biocompatibility of FOE treatment. Collectively, the localized fluorescence change is indicative of the sub-millimeter spatial resolution of the FOE, holding promise for broad applications including drug delivery and neuron stimulation for minimized side effect, as well as localized gene transfection for gene-protein studies.

**Conclusion**

Fiber based photoacoustic emitters composed of nanoparticle-polymer matrix with superior optical and mechanical properties were designed and fabricated. The new two coating design, including a diffusion layer and an expansion layer, provides more controllability in amplitude and frequency of the ultrasound generated. Localized acoustic wave generation with high amplitude and tunable frequency in the sub MHz range were achieved. By characterizing the optoacoustic signal profile in amplitude and frequency spectrum, a matrix of CNTs/PDMS was demonstrated to be a preferable candidate to achieve high amplitudes. Two effective strategies to control the acoustic frequency were demonstrated. First, the frequency was varied by the thickness of the diffusion layer. Second, changing the absorber concentration also successfully achieved tunable frequency. A sub-MHz frequency acoustic with sub-millimeter confinement was produced using the miniaturized FOE, overcoming the limitation of other existing ultrasound sources. Such FOE device design holds promise for a wide range of cellular applications, including cell membrane sonoporation, and offers new tools for localized drug delivery, neuron stimulation and gene transfection with high efficacy and less side effect.

By achieving the high miniaturization levels demonstrated, the tunable optoacoustic emitters are promising for minimally invasive medical applications, where the fiber based optoacoustic devices presented here could be inserted in syringe needles or catheters in close proximity to a focal lesion, thus overcoming the problem of reduced precision and amplitude induced by traditional focused ultrasound. Under a broader context, this technique offers the potential to generate stable and reversible sonoporation at each focal target through modification of the ultrasound parameters, enabling precise control for biomedical ultrasound application, which is not available with existing technologies, especially for drug delivery and gene transfer. Additionally, this FOE is immune to electromagnetic interference and hence is compatible with magnetic resonance imaging (MRI)[28]. Flexibility, along with its unprecedentedly miniaturization, and amenability to be readily repeated, make it a potentially transformative technology.

**Data and materials availability**

The authors declare that all of the data supporting the findings of this study are available within the paper and the supplementary information.

**Acknowledgements**

This research was supported by R01 NS109794 to JXC and Boston University Nanotechnology Innovation Center Pilot Grant (2018-2019) to CY.

**Conflict of interests**

The authors declare no conflict of interest.

**Contributions**

LS, LL, CY and J-XC conceived this project. LS, YZ and LL designed the FOE system, LS and YZ built the characterization set up. LS and YH performed the experiments. LS, YJ and YH implemented the sonoporation experiment. LS, J-XC and CY wrote the manuscript. CY and J-XC provided overall guidance for the project.

**Figures**

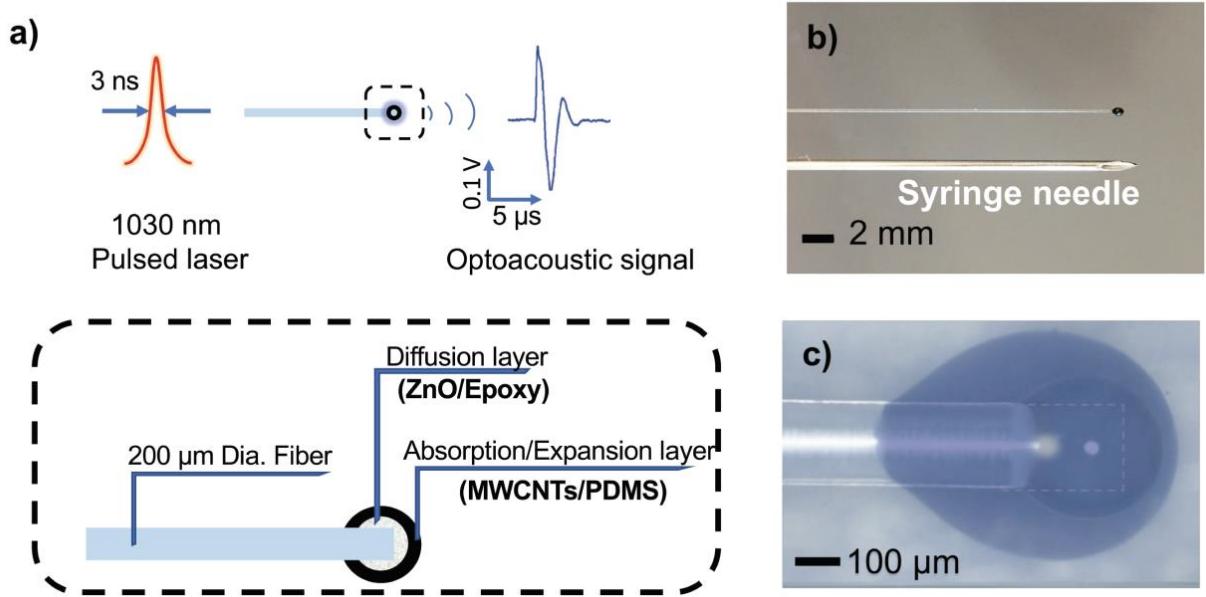

**Figure 1. Design and fabrication of a two-layer fiber-based optoacoustic emitter (FOE). a)** Schematic of optoacoustic effect and the design of FOE with a two-layer structure. **b)** Comparison between the fiber-based emitter and a syringe needle (20G, ID 0.6mm, OD 0.91mm). **c)** Micrographs of a fabricated emitter. The image transparency was adjusted to visualize the inner diffusion layer and the outer absorption/expansion layer. White dash line outlines the fiber distal end.

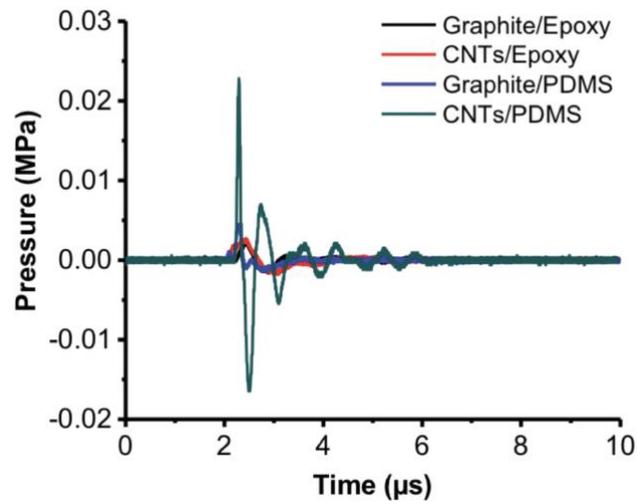

**Figure 2. Comparison of optoacoustic signal based on different materials.** Pressure amplitude in time domain using Epoxy/Graphite (black), Epoxy/CNTs (red), PDMS/Graphite (blue) and PDMS/CNTs (green) respectively.

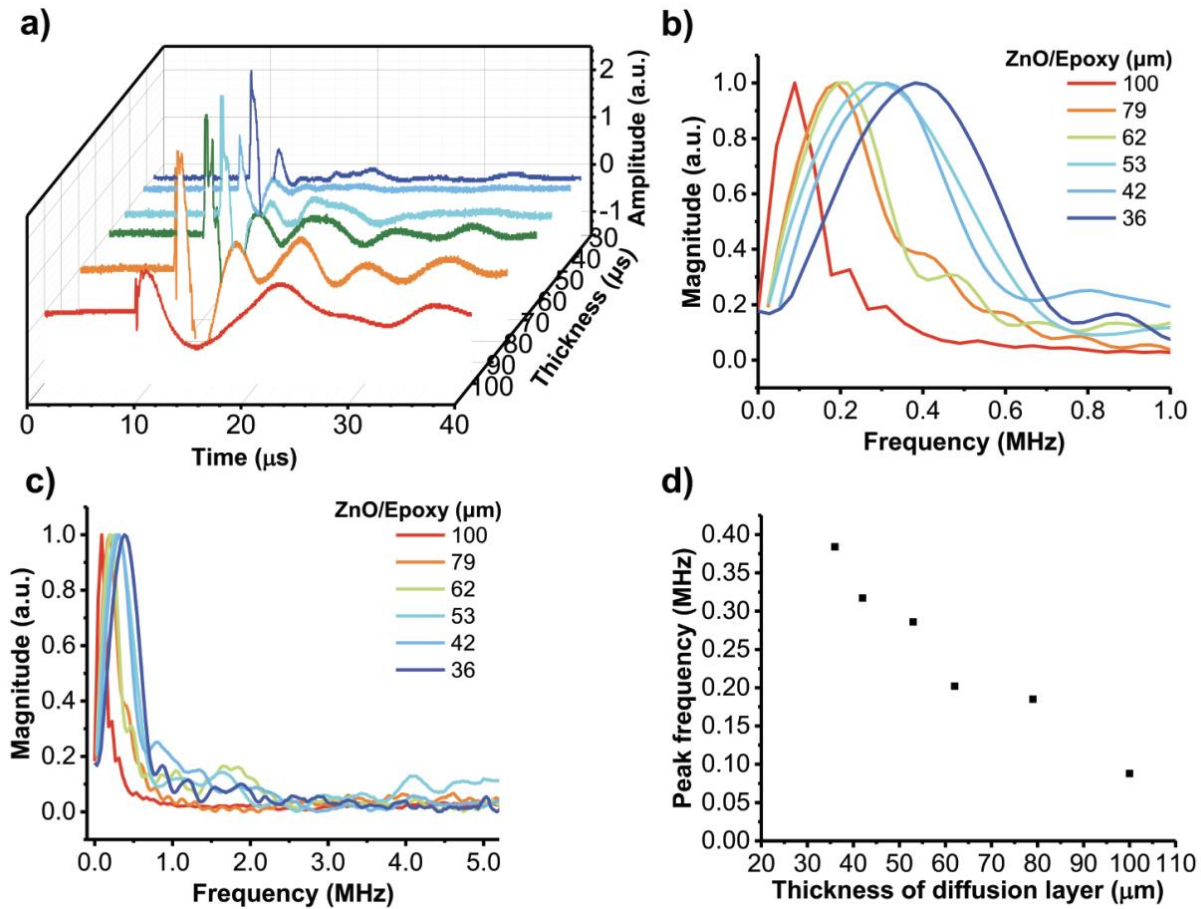

**Figure 3. Controlling the ultrasound frequency via modifying the diffusion layer. a)** The ultrasound signals in time domain from FOEs fabricated with diffusion layers (ZnO/Epoxy) of 36-100 μm and absorption/thermal expansion layer (CNTs/PDMS) of 109±24 μm. **b)** The frequency domain within 0-1.0 MHz of the ultrasound. **c)** Zoomed-out figure with frequency ranging from 0 to 5.0 MHz. **d)** Ultrasound peak frequency plotted as a function of the diffusion layer thickness.

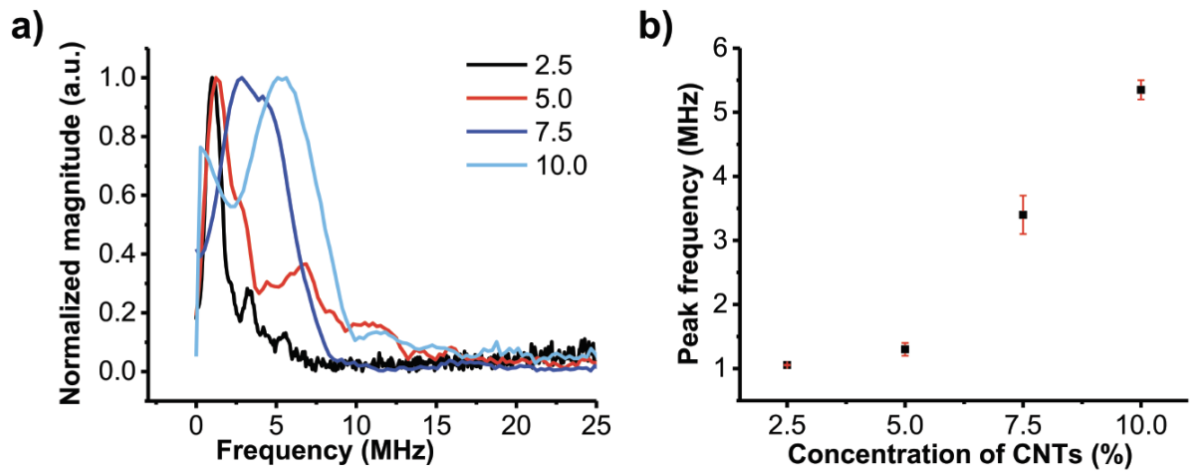

**Figure 4. Optoacoustic signal frequency as a function of effective absorption/expansion layer thickness. a)** Normalized frequency spectrum for FOEs with a different thermal expansion layer (CNTs/PDMS, 2.5%, 5.0%, 7.5%, 10.0% by weight). **b)** Peak frequency plotted as a function of CNTs/PDMS concentration. Each data point in **b)** is the average value of two identical FOEs for each concentration and error bars are the standard deviation.

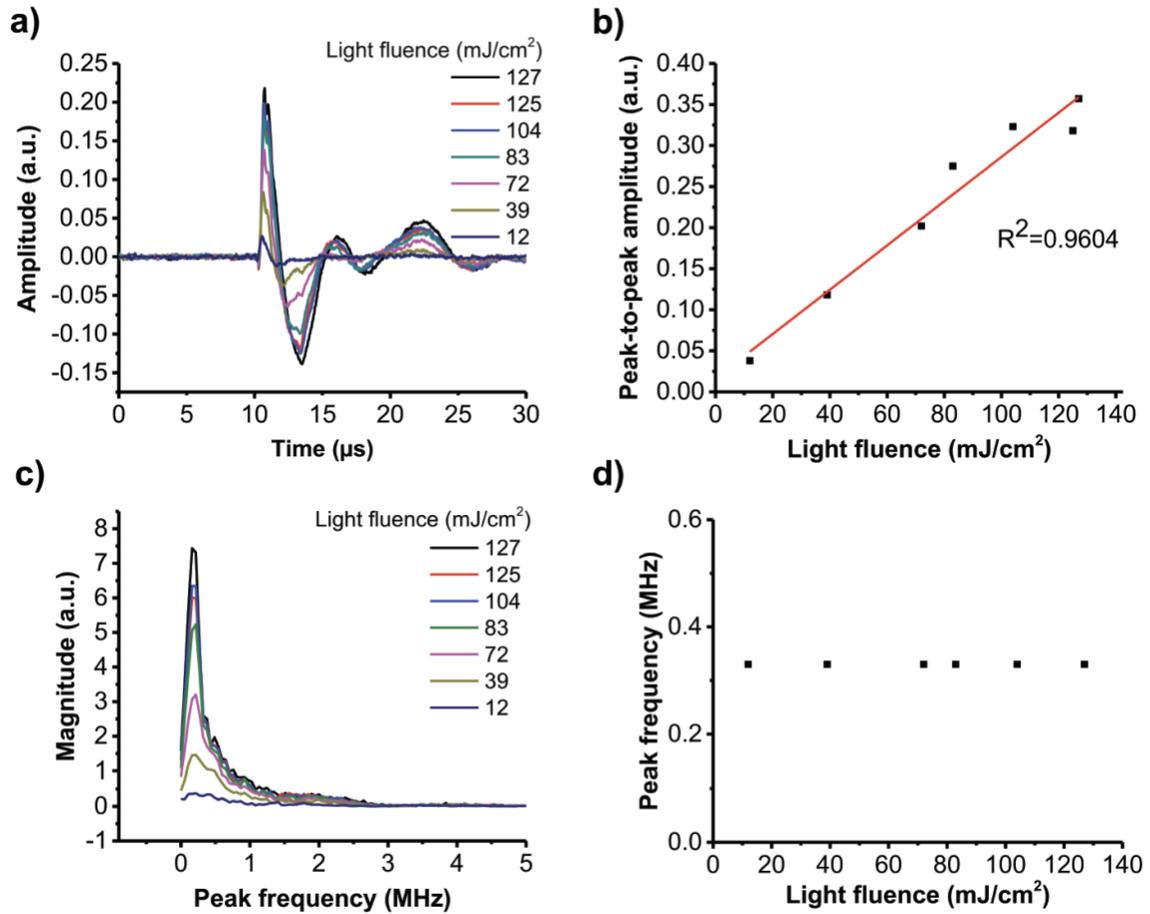

**Figure 5. Acoustic amplitude and frequency as functions of laser pulse energy. a)** Acoustic amplitude measured in time domain when varying laser pulse energy from 12 to 127 mJ/cm2 using optical attenuators. **b)** Peak-to-Peak amplitude of acoustic as a function of laser pulse energy. Solid line: fitting of the data. **c)** Acoustic signal in frequency domain. **d)** Acoustic peak frequency as a function of laser pulse energy.

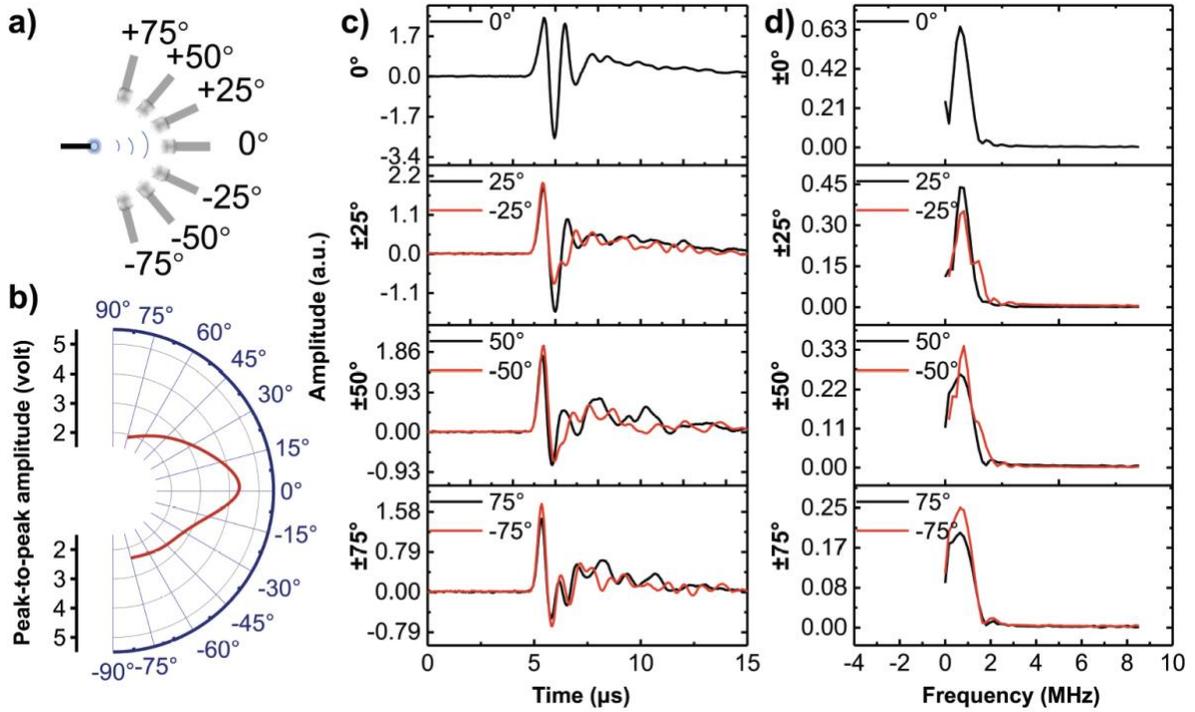

**Figure 6. Characterization of acoustic angular distribution. a)** Schematic of the detection. **b)** Acoustic peak to peak amplitude detected at angles 0°, ±25°, ±50°, and ±75°. **c)** Angle dependence of acoustic peak-to-peak amplitude. **d)** Frequency spectra for acoustic signals detected at these angles.

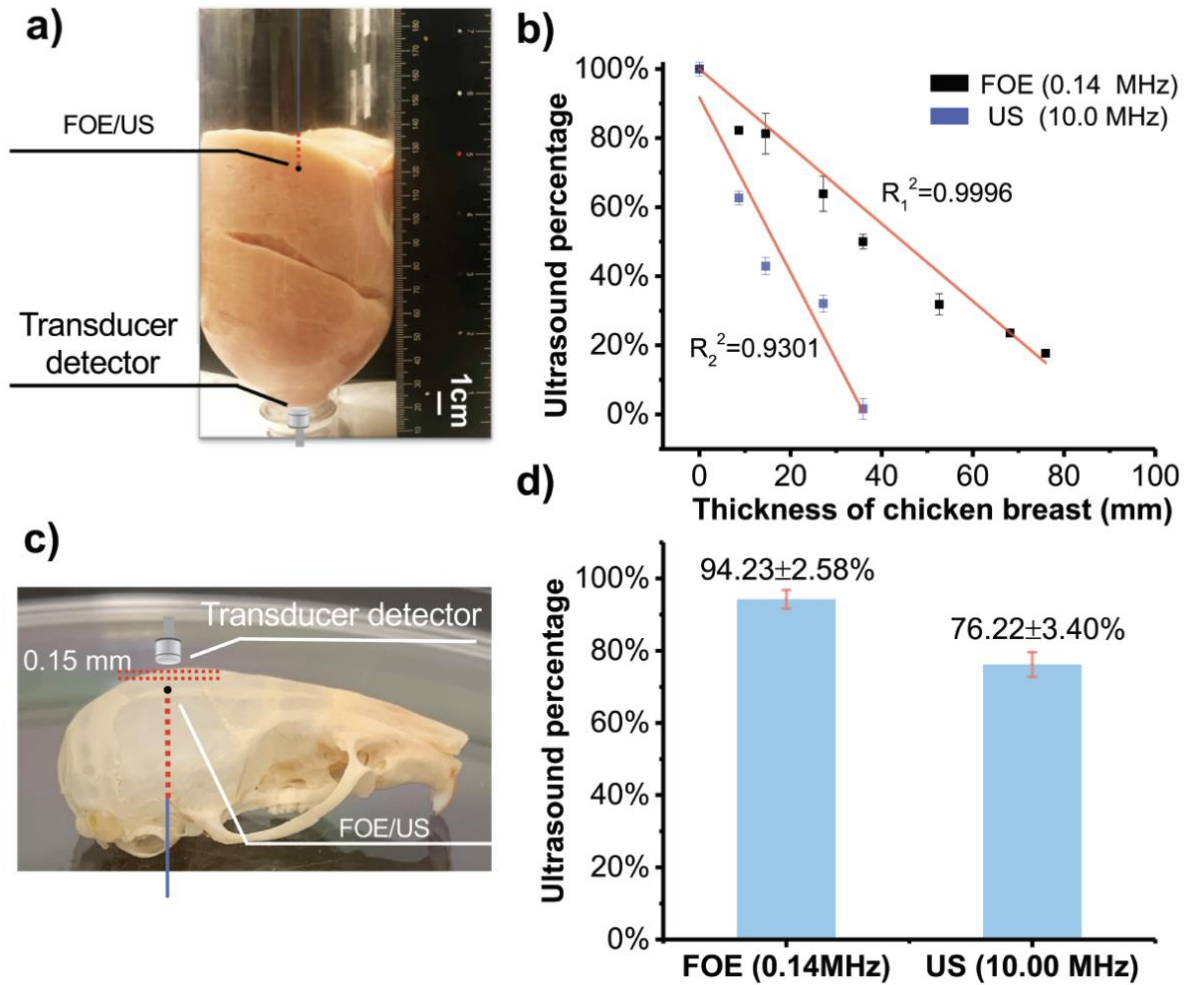

**Figure 7. Attenuation of low frequency acoustic signals generated by the fiber converter in chicken breast and a mouse skull.** A fiber-based FOE and a transducer with peak frequency of 10 MHz (transmit mode) were used as acoustic sources. A transducer with peak frequency of 5 MHz (receive mode) was used as the detector. Set up for the acoustic attenuation in chicken breast **a)** and a mouse skull **c)**. **b)** Acoustic amplitude percentage after penetration in different thickness of chicken breast for acoustic signals generated by the FOE (black) and the transducer in the transmit mode (blue). Error bar was obtained by detection at 3 different locations in chicken breast. **d)** Acoustic amplitude after penetrating a mouse skull. Error bar was obtained by detection at 3 different locations in the mouse skull.

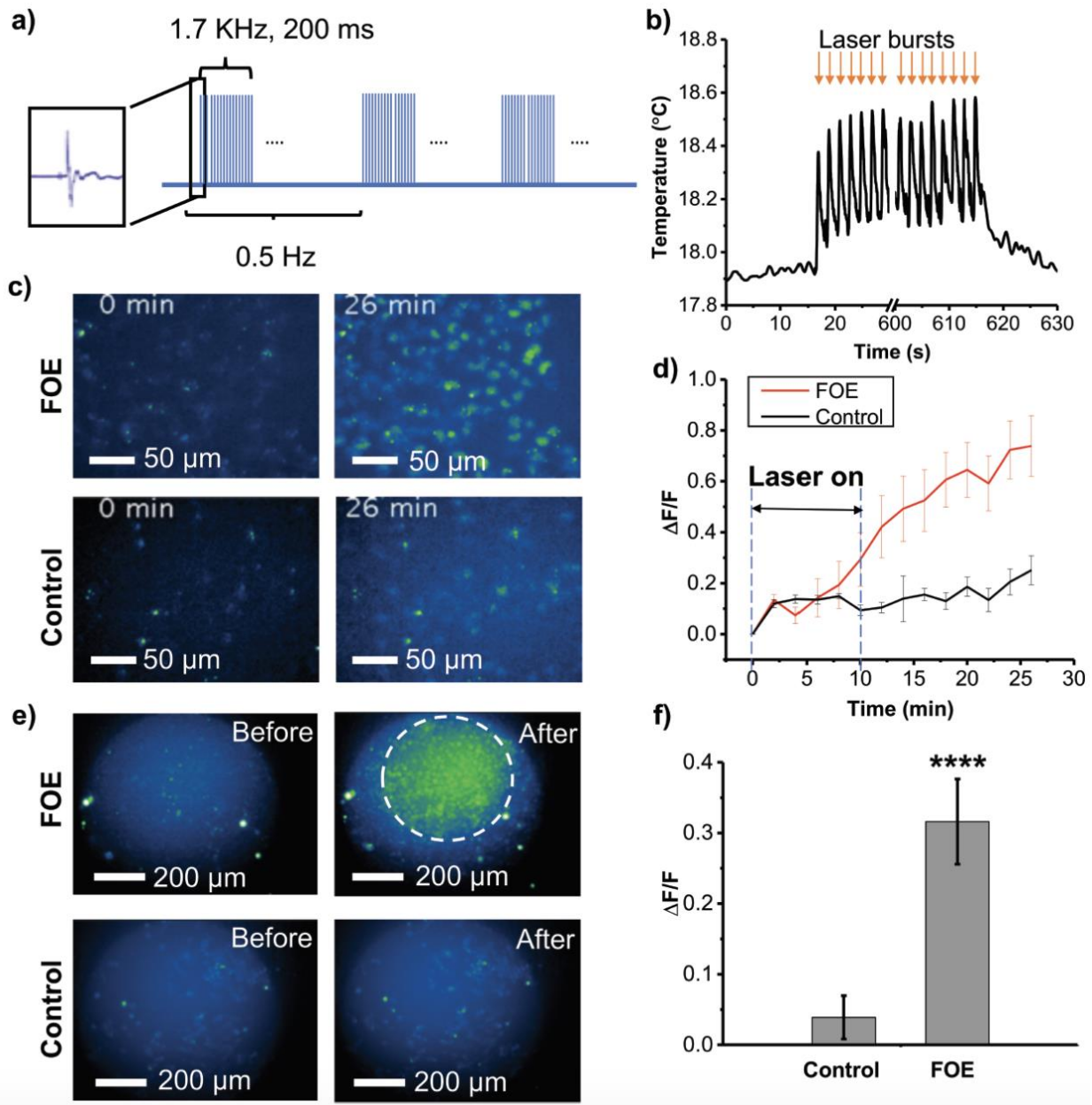

**Figure 8. The low-frequency ultrasound from the FOE facilitates Sytox cellular uptake. a)** Schematic of laser pulse tone bursts. **b)** Temperature change at the FOE tip during FOE treatment. Laser is on at 10 s and continued for 8 min. The signal is smoothened while keeping the overall temperature change constant. **c)** Fluorescence imaging of FOE treated region at 0 min and 26 min compared to the control group, images were taken with 20 × objective. **d)** Averaged fluorescence intensity changing dynamics of 30 cells in the treated region. Red: FOE treated cells. Black: control

group. **e)** Fluorescence imaging of FOE treated group and control group taken with 10× objective. The white dash circle indicates the region of significant fluorescence intensity change observed, which was right beneath the position of the FOE at a distance of 100 μm above the cells. **f)** Comparison of fluorescence intensity change between FOE treated group and control group. **** $P < 0.0001$.